\documentclass[onecollarge,natbib]{svjour2}
\bibpunct{[}{]}{;}{n}{}{,}
\smartqed
\usepackage{graphicx}
\usepackage{mathptmx}
\journalname{Few-Body Systems}
\begin{document}

\title{On $J^{PC}=0^{--}$ exotic glueball
}

\author{Loredana Bellantuono
}

\institute{L. Bellantuono \at
              Dipartimento di Fisica, Universit\`a di Bari, and INFN, Sezione di Bari, via Orabona 4, I-70126 Bari, Italy\\
							\email{loredana.bellantuono@ba.infn.it}
}

\date{Received: date / Accepted: date}

\maketitle

\begin{abstract}
The mass spectrum of the gluonium with $J^{PC}=0^{--}$ is examined in three bottom-up AdS/QCD models. The results are used to identify several production and decay modes useful for searching this state. Moreover, the properties of such glueball in a hot and dense quark medium are discussed.
\end{abstract}

\section{Introduction}
\label{intro}
Glueballs, bound states of gluons arising from the non-Abelian nature of strong interactions, are an important testbed for non perturbative aspects of QCD. The main obstacle in their search is the mixing with quarkonia $(\bar{q}q)$ with the same quantum numbers. A promising strategy for their identification is to focus on exotic states whose $J^{PC}$ quantum numbers are unaccessible to quark-antiquark configurations. This is the case of several gluonia with negative $C$-parity, composed of an odd number of constituent gluons (``oddballs"), for which little theoretical information is available.
In particular, for $J^{PC}=0^{--}$, the mass predictions for the lightest state span the range from $m_{0^{--}}=2.79$ GeV in the flux-tube model \cite{Isgur}, to $m_{0^{--}}\approx 5.166$ GeV in lattice QCD simulations \cite{Gregory}. Two stable $0^{--}$ oddballs, with masses $m_{0^{--}}=3.81 \pm 0.12$ GeV and $m_{0^{--}}=4.33 \pm 0.13$ GeV, have been predicted by QCD sum rules \cite{Qiao2014}. 

The mass spectrum of the $J^{PC}=0^{--}$ oddball can be computed in a framework inspired by the AdS/CFT correspondence. This duality conjecture relates a strongly coupled gauge theory in a four dimensional $(4D)$ Minkowski space to a semiclassical gravity theory defined in a five dimensional $(5D)$ anti-de Sitter (AdS) geometry times a $5D$ sphere \cite{Maldacena}. In Poincar\'e coordinates, the line element
\begin{equation}
\label{AdS5} 
ds^2=\frac{R^2}{z^2}(dx_{0}^2-d\vec{x}^2-dz^2) \qquad\qquad z>0 \,,
\end{equation}
with $z$ the fifth holographic coordinate, describes the AdS bulk metric. The original formulation of gauge/gravity duality required the $4D$ theory to be conformal invariant. Holographic bottom-up models, constructed to reproduce QCD properties, break such a symmetry by introducing an infrared energy scale in the bulk; the $J^{PC}=0^{--}$ oddball mass spectrum can then be determined either in vacuum or in a quark bath at finite temperature and density. Such an approach complements the top-down methods applied to analyze, e.g., the scalar $0^{++}$ gluonium \cite{Csaki,Brunner2015PRL}.
\section{$J^{PC}=0^{--}$ mass spectrum in three holographic models of QCD}
\label{sec:1}
The holographic correspondence is based on a dictionary, according to which a local gauge-invariant operator in the $4D$ field theory is dual to a field in $5D$ AdS \cite{Witten,Gubser}. In QCD, an interpolating current representing the glueball with quantum numbers $J^{PC}=0^{--}$ can be written in terms of the gluon field strengths $G^a_{\mu \nu}(x)$ and $\widetilde G^a_{\mu \nu}(x)=\frac{1}{2} \epsilon_{\mu \nu \rho \sigma}G^a_{\rho \sigma}(x)$, namely
\begin{equation}
\label{J0}
J_{0}(x)=g_s^3 d_{abc} [\eta^t_{\alpha \beta} \widetilde G^a_{\mu \nu}(x)] [\partial_{\alpha}  \partial_{\beta} G^b_{\nu \rho}(x)] [G^c_{\rho \mu}(x)]\,,
\end{equation}
with $a,b,c$ color indices, $d_{abc}$ the symmetric tensor defining the anticommutator of $SU(3)_{c}$ generators, and $g_{s}$ the strong coupling constant.
The transverse $\eta^t_{\alpha \beta}$ metric is defined as $\eta^t_{\alpha \beta}= \eta_{\alpha \beta} - \frac{\partial_\alpha \partial_\beta}{\partial^2}$, with $\alpha,\beta$ (as well as $\mu,\nu$) $4D$ Lorentz indices, and $\eta_{\alpha \beta}$ the Minkowski metric tensor \cite{Qiao2014,Qiao2015}. The operator (\ref{J0}) has conformal dimension $\Delta=8$, and its holographic dual field, $O_{0}(x,z)$, has mass obtained by the relation $M_{5}^{2}R^{2}=\Delta(\Delta-4)$ \cite{Witten,Gubser}. In the following, the $AdS_{5}$ radius is set to $R=1$.
The $5D$ action for $O_{0}(x,z)$ can be written as
\begin{equation}
\label{action}
S=\frac{1}{k} \int d^5 x  \sqrt g a(z)\left[ g^{MN} \partial_M O_0\,  \partial_N O_0 -M_5^2 O_0^2 \right] \,,
\end{equation}
where $g_{MN}$ is the bulk metric, $g=|det (g_{MN}) |$ and $k$ is a parameter making the action dimensionless. 
To account for confinement in QCD, conformal invariance must be broken in the action (\ref{action}). Three different models including such effect, either by a proper choice of the function $a(z)$ or by introducing a dynamical field in the metric, are discussed in the following. The mass spectrum can be determined solving the Euler-Lagrange equations for the field $O_{0}(x,z)$.
\paragraph{Hard-wall model.} 
A simple way of modeling confinement in the holographic setup is by considering a slice of the $AdS_{5}$ space, with a sharp cutoff at a finite distance $z_{m}$ along the fifth dimension \cite{Erlich}. The metric $g_{MN}$ is given by $(\ref{AdS5})$, and the condition $z\leq z_{m}$ is implemented in the action through $a(z)=\Theta\left(z_{m}-z\right)$, with $\Theta$ the Heaviside function. The cutoff $z_{m}$ sets a mass scale. The choice $1/z_{m}=346$ MeV, obtained from analyses on axial and vector mesons \cite{Erlich}, gives the results $m_{0}=2.80$ GeV and $m_{1}=4.14$ GeV for the lowest-lying and the first-excited $0^{--}$ oddball, respectively \cite{Bellantuono2015}.
\paragraph{Soft-wall model.}
In this framework, the geometry is $AdS_{5}$ (\ref{AdS5}), and conformal invariance is smoothly broken by the function $a(z)=e^{-c^{2}z^{2}}$ which introduces a mass scale $c$ in the action \cite{Karch}. The Regge-like mass spectrum is obtained \cite{Bellantuono2015}: 
\begin{equation}
\label{mass spectrum SW}
m_{n}^{2}=4c^{2}(n+4)\,.
\end{equation}
Setting $c=m_{\rho}/2=388$ MeV from the $\rho$ meson mass computed in the model \cite{Karch}, one obtains $m_{0}=1.55$ GeV and $m_{1}=1.74$ GeV \cite{Bellantuono2015}. The mass spectrum (\ref{mass spectrum SW}) corresponds to the poles of the two-point correlation function of $J_{0}(x)$,
\begin{equation}
\label{qcd-corrfun}
\Pi(p^2)= i \int d^4 x \, e^{i p x} \langle 0| T[J_0(x) J^\dagger_0(0)]  | 0 \rangle \,.  
\end{equation}
In the AdS/CFT dictionary, the QCD interpolating current $J_{0}(x)$ is interpreted as the source of the dual field $O_{0}(x,z)$. The $4D$ Fourier transforms of such operators, $\tilde{J}_{0}(p)$ and $\tilde{O}_{0}(p,z)$, are related by the bulk-to-boundary propagator $\tilde{K}(p,z)$ of the oddball field, through $\tilde{O}_{0}(p,z)=\tilde{K}(p,z)\tilde{J}_{0}(p)$. Holography prescribes the identification between the partition functions of the dual theories, and this allows to compute the two-point correlation function (\ref{qcd-corrfun}) as
\begin{equation}
\Pi\left(p^{2}\right)\approx\left.\frac{\delta^{2}S_{os}}{\delta \tilde{J}_{0} \delta \tilde{J}_{0}}\right|_{\tilde{J}_{0}=0}\,,
\end{equation}
with $S_{os}$ the $5D$ on-shell action. The poles of $\Pi\left(p^{2}\right)$ are given in (\ref{mass spectrum SW}).
\paragraph{Einstein-dilaton model.}
A third possibility is to consider a class of dynamical bottom-up models reproducing confinement through a distorsion of the bulk geometry \cite{Li,Chen}:
\begin{equation}
\label{metric-ED}
ds_{(ED)}^2=\frac{e^{2 A_s(z)-\frac{4}{3} \Phi(z)}}{z^2} \left[dx_0^2-d\vec x^2 -dz^2  \right] \,.  
\end{equation}
$\Phi(z)$, a dilaton field, couples to the graviton, and its profile is determined solving the Einstein equations deduced from this metric. The function $A_{s}(z)$ can be chosen as $A_{s}(z)=\delta^2 z^{2}$, with the mass scale fixed to $\delta=0.43$ GeV \cite{Li}. Once the profile of $\Phi(z)$ is obtained, the two lightest oddball states turn out to have mass $m_{0}=2.82$ GeV and $m_{1}=4.07$ GeV \cite{Bellantuono2015}.
\paragraph{}With these results, several production and decay modes can be identified, useful for the search of the $0^{--}$ gluonium. They are reported in Table \ref{tab:1} for the specific case $m_{0^{--}}=2.8$ GeV \cite{Bellantuono2015}.
\begin{table}[t]
\caption{Production and decay modes of the $J^{PC}=0^{--}$ oddball, for $m_{0^{--}}=2.8$ GeV.}
\centering
\label{tab:1}
\begin{tabular}{lll}
\hline\noalign{\smallskip}
Radiative production & Hadronic production & Decay mode  \\[3pt]
\tableheadseprule\noalign{\smallskip}
$\chi_{c1}(3510) \to \gamma \, G(0^{--})$ & $X(3872) \to \omega \, G(0^{--})$ & $ G(0^{--}) \to \gamma \, f_1(1285)$ \\
$X(3872) \to \gamma \, G(0^{--})$ & $h_{c}(3525) \to \pi \pi \, (I=0)  \, G(0^{--})$ & $ G(0^{--}) \to \omega \, f_1(1285)$ \\
$\chi_{c2}(3556) \to \gamma \, G(0^{--})$ & $\chi_{b1}(10255) \to (\omega, \phi, J/\Psi)  \, G(0^{--})$ & $ G(0^{--}) \to \rho \, a_1(1260)\,\,\,(I=0)$ \\
$\chi_{c2}(3927) \to \gamma \, G(0^{--})$ & $\Upsilon(nS) \to (f_1(1285), \chi_{c1}, X(3872))  \, G(0^{--})$\\ & \, \\
$\chi_{b1}(9892) \to \gamma \, G(0^{--})$ & $h_{b}(9899) \to f_0(980) \, G(0^{--})$ & $ G(0^{--}) \to h_1(1270) \, f_0(980)$ \\
$\chi_{b1}(10255) \to \gamma \, G(0^{--})$ & $h_{b}(10260) \to f_0(980) \, G(0^{--})$ & $ G(0^{--}) \to \rho \, \pi \,\,\,(I=0) $ \\
$\chi_{b2}(9912) \to \gamma \, G(0^{--})$ & $h_{b}(9899) \to G(0^{++}) \, G(0^{--})$ & $ G(0^{--}) \to K^* \, K \,\,\,(I=0) $ \\
$\chi_{b2}(10269) \to \gamma \, G(0^{--})$ & $h_{b}(10260) \to G(0^{++}) \, G(0^{--})$ & $ G(0^{--}) \to (\eta,\eta^\prime)(\omega, \phi) $ \\
\noalign{\smallskip}\hline
\end{tabular}
\end{table}
\section{Oddball in thermalized and dense medium}
\label{sec:2}
\begin{figure}
\centering
  \includegraphics[width=0.47\textwidth]{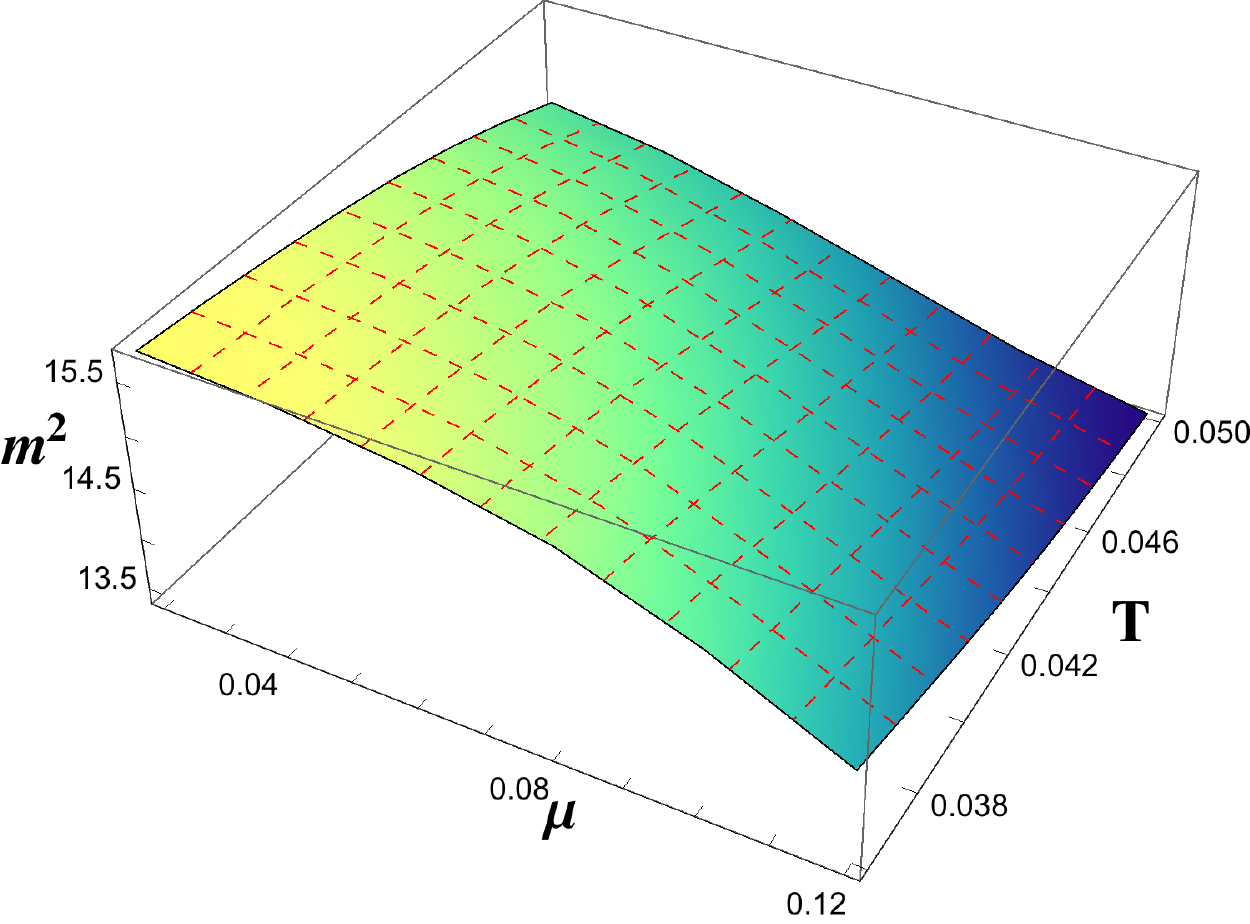} \hspace{0.6cm}
	\includegraphics[width=0.47\textwidth]{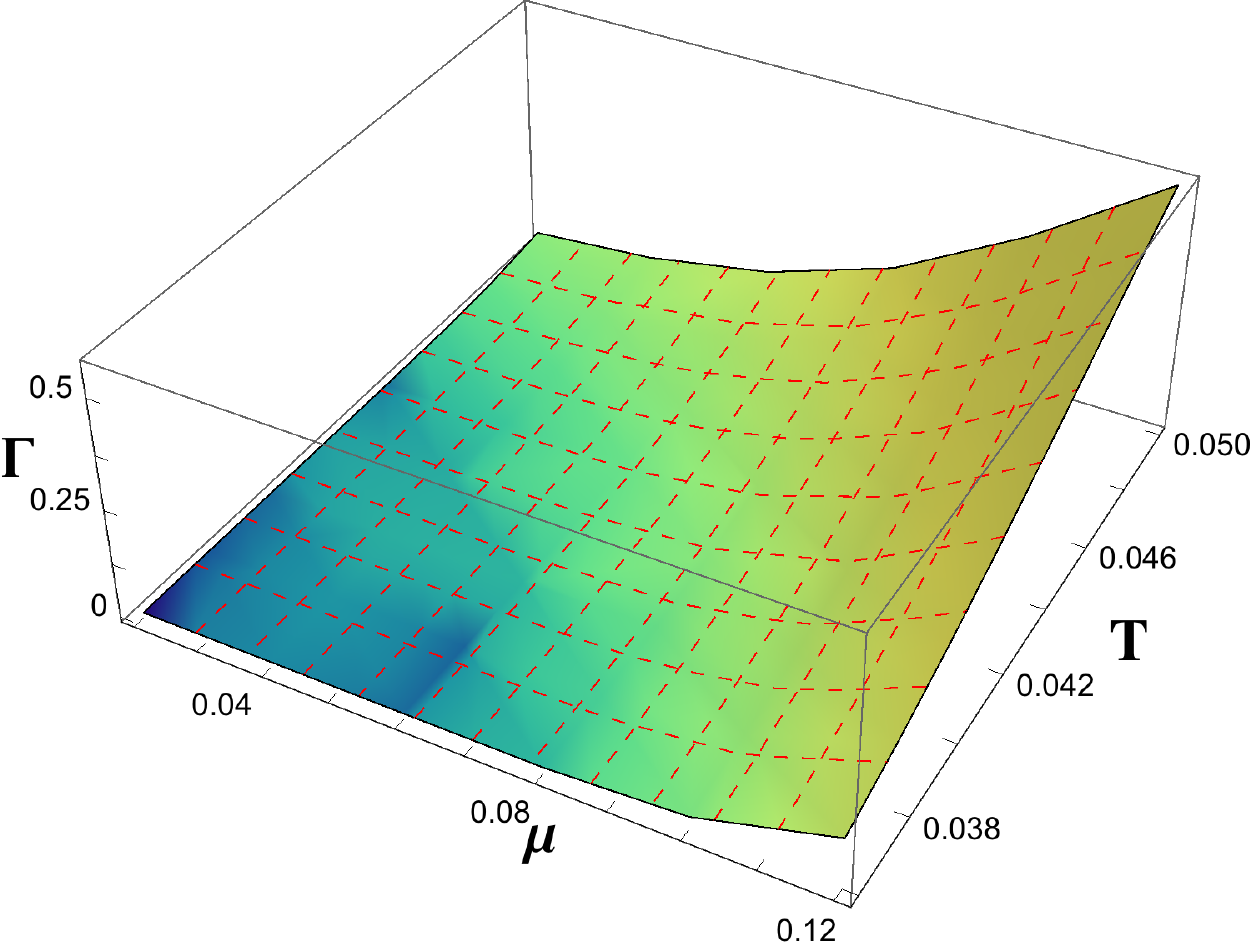}
\caption{Squared mass $m^{2}$ and width $\Gamma$ of the lightest $J^{PC}=0^{--}$ glueball, obtained from the soft-wall model with AdS/RN metric, in a range of temperature $T$ and chemical potential $\mu$, in units of $c$ \cite{Bellantuono2015}.}    
\label{fig:3D}
\end{figure}

The AdS/QCD duality allows to examine the properties of a $J^{PC}=0^{--}$ oddball in a quark thermal bath, at finite temperature $T$ and chemical potential $\mu$. Stability of the gluon configuration against thermal and density fluctuations can be investigated in comparison with other quark and gluon bound states. In-medium effects can be incorporated in the holographic description using the action (\ref{action}), with an appropriate bulk geometry. I explicitly discuss the soft-wall case. 
The $5D$ line element
\begin{equation}
\label{metric-medium}
ds^2= \frac{1}{z^2}\left( f(z) dx_0^2-d\vec x^2- \frac{dz^2}{f(z)} \right)  
\end{equation}
is characterized by a metric with $f(z)$ which can be related to temperature and density of the boundary theory.
The $5D$ Reissner-Nordstr\"om $AdS$ geometry (AdS/RN), with
\begin{equation}
\label{RN-f}
f(z)=1- \left( \frac{1}{z_h^4} +q^2 z_h^2 \right) z^4+q^2 z^6  \,,
\end{equation}
can be used as the holograpic dual of a thermalized and dense medium in the deconfined phase \cite{Colangelo2012,Lee}. This charged black hole bulk metric has an outer horizon $z=z_{h}$ and charge $q$ which can be related to the temperature $T$ and chemical potential $\mu$ of the dual boundary theory. Defining $Q=q z_h^3$ and imposing the condition  $0\leq Q\leq \sqrt{2}$, the  black-hole temperature is
\begin{equation}
\label{T}
 T=\frac{1}{4\pi}\left| \frac{df}{dz}\right|_{z=z_h}=\frac{1}{\pi z_h} \left(1-\frac{Q^2}{2} \right) \,.
\end{equation}
In the QCD generating functional, $\mu$ multiplies the quark number operator $O_{q}(x)=q^{\dagger}(x)q(x)$. Thus, it can be interpreted, according to gauge/gravity correspondence, as the source of a bulk field dual to $O_{q}(x)$, the time component of a $U(1)$ gauge field $A_{M}(x,z)$. By rotational invariance, the spatial components $A_{i}$ (with $i=1,2,3,z$) vanish, while $A_{0}\approx \mu-\kappa q z^{2}$ for small $z$. The condition $A_{0}(z_{h})=0$ provides a linear relation between $\mu$ and $q$:
\begin{equation}
\label{mu}
\mu=\kappa \frac{Q}{z_{h}}\,,
\end{equation}
with $\kappa$ a dimensionless parameter that will be fixed to $1$, a choice determining the quark chemical potential $\mu$ up to a numerical factor. 
\begin{figure}
\centering
  \includegraphics[width=0.47\textwidth]{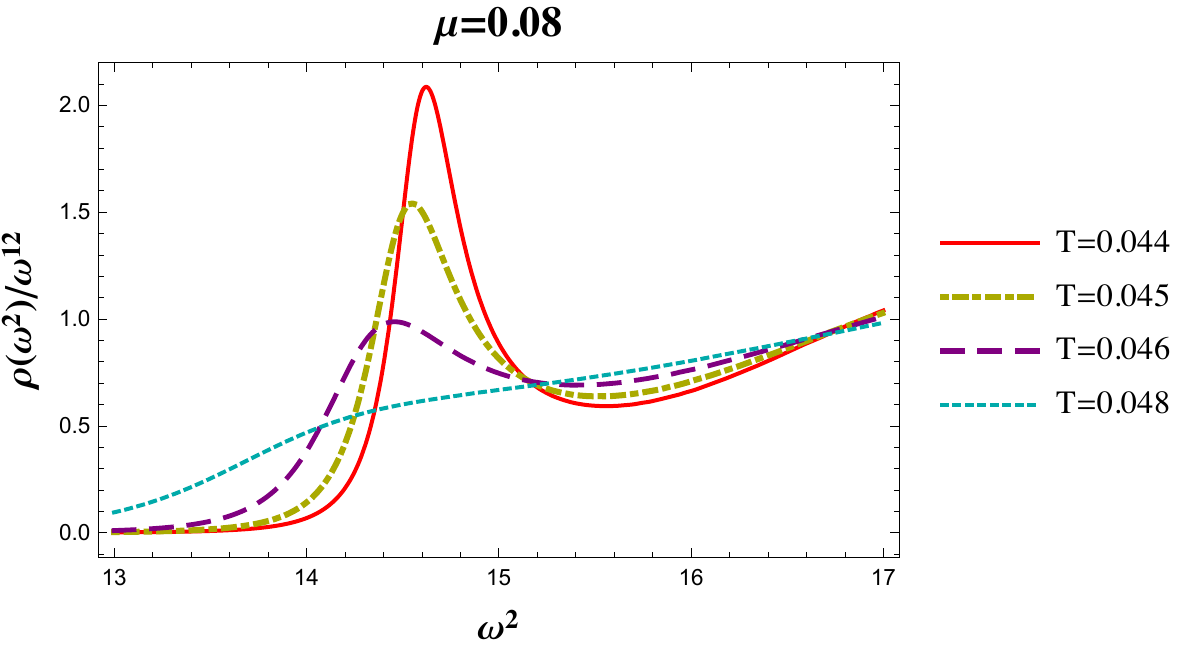} \hspace{0.6cm}
	\includegraphics[width=0.47\textwidth]{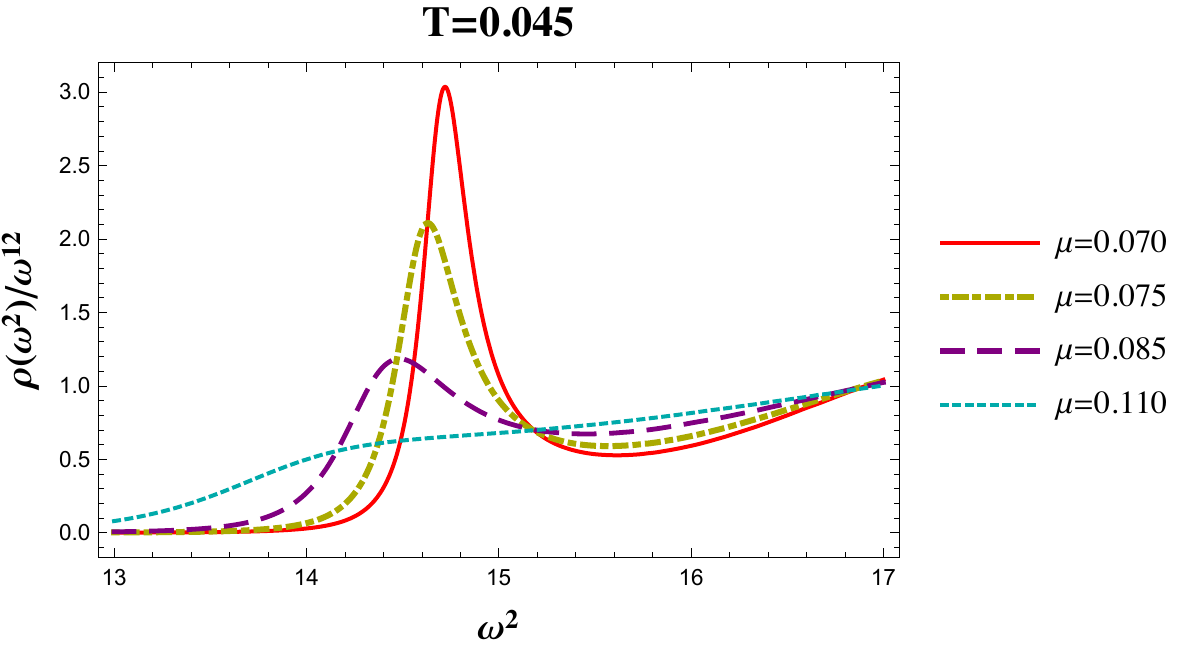}
\caption{Spectral function $\rho\left(\omega^{2}\right)/\left(\omega^{2}\right)^{6}$ computed in the soft-wall model with AdS/RN metric, at fixed $\mu$ (left) or $T$ (right). The dimensionful quantities are in units of $c$, and the constant $10^{-8}k/\left(2z_{h}^{8}\right)$ has been factorized out in the spectral function \cite{Bellantuono2015}.}
\label{fig:rho}     
\end{figure}
Important information on the QCD bound states in a quark thermal bath can be inferred from the spectral function $\rho\left(\omega^{2}\right)$, the imaginary part of the retarded Green function. The spectral function of the $0^{--}$ oddball is represented in vacuum by an infinite number of delta functions centered at the eigenvalues of the mass spectrum (\ref{mass spectrum SW}). 
Fig. \ref{fig:rho} shows the changes of the in-medium $\rho\left(\omega^{2}\right)$ as temperature and chemical potential are switched on. At any finite and fixed value of $\mu$, the peaks of the spectral function broaden and move towards lower values of $\omega^{2}$ when $T$ increases. An analogous behaviour is observed by keeping the temperature fixed, and increasing the chemical potential. The $(T,\mu)$-dependence of the lightest oddball's squared mass and width is shown in Fig. \ref{fig:3D}. The values of the temperature and chemical potential at which the lowest-energy peak becomes indistinguishable in the profile of the spectral function, are smaller than those obtained for light vector \cite{Fujita,Giannuzzi}, scalar $\bar{q}q$ mesons \cite{Colangelo2008}, the lightest scalar glueball \cite{Colangelo2007}, and hybrid mesons \cite{Bellantuono2014}. Hence, $0^{--}$ oddballs are found to suffer of larger in-medium instabilities with respect to other hadrons.

\begin{figure}
\centering
  \includegraphics[width=0.47\textwidth]{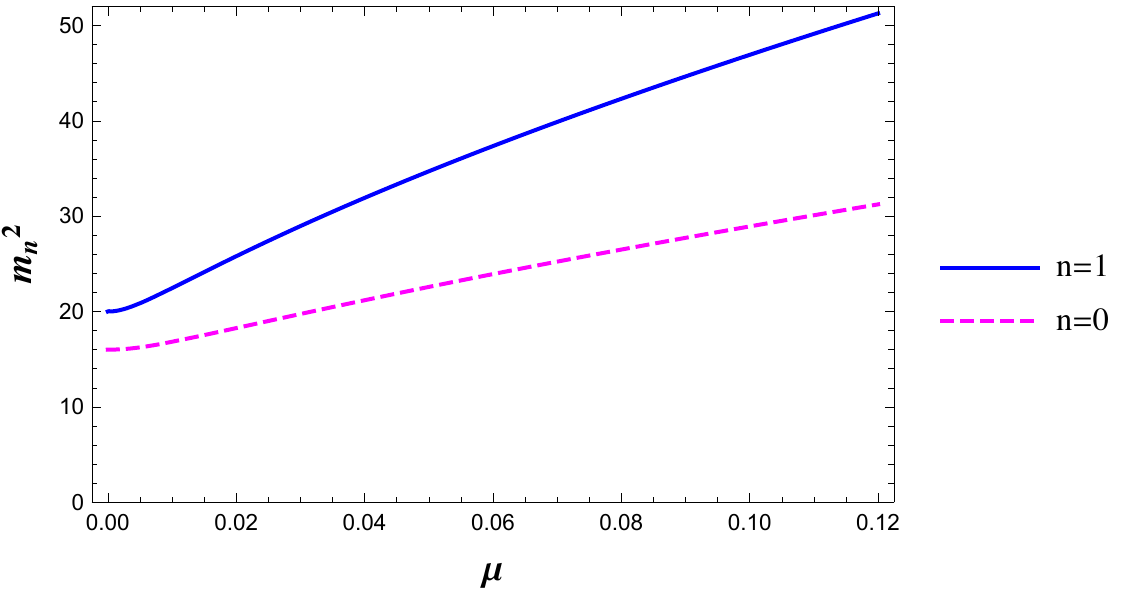}
\caption{Squared mass $m^{2}$ as a function of $\mu$ for the two lightest $J^{PC}=0^{--}$ oddballs, in the tcAdS soft-wall model.}
\label{fig:tcAdS}
\end{figure}

For points in the $T-\mu$ plane below the QCD deconfinement transition, the state described by AdS/RN metric is metastable \cite{Lee}. The thermal-charged AdS metric (tcAdS), with line element (\ref{metric-medium}) and deformation
\begin{equation}
\label{tcAdS-f}
f(z)=1+q^{2}z^{6}\,.
\end{equation}
is proposed as a dual of the confined phase of QCD at small temperature and finite chemical potential \cite{Lee,Park}. This geometry is related to the AdS/RN one by the Hawking-Page transition, which describes deconfinement in the holographic framework. 
The chemical potential, related to the finite density of the hadronic medium, grows linearly with $q$, while the temperature is implemented through a periodicity in the Euclidean time coordinate $\tau=ix_{0}$ \cite{Park}. The results, depicted in Fig. \ref{fig:tcAdS}, reveal that the masses of the two lightest oddballs increase with $\mu$, a different behaviour with respect to what is found in the deconfined phase \cite{Bellantuono2015}.
\section{Conclusions}
\label{conclusions}
In AdS/QCD, the mass of the lowest-lying $0^{--}$ oddball is found to be lighter than as computed by other approaches.
The effect of deconfined quark matter can be modeled by the AdS/RN geometry: oddballs are found to suffer from larger thermal and density instabilities with respect to other hadrons. On the other hand, the holographic description of a confined medium, obtained using the tcAdS metric, gives a mass which increases with the chemical potential.

\begin{acknowledgements}
I thank P. Colangelo and F. Giannuzzi for collaboration, and F. De Fazio and S. Nicotri for discussions.
\end{acknowledgements}

\end{document}